\documentclass[aps,prl,reprint,groupedaddress]{revtex4-1}

\usepackage{graphicx}
\usepackage{epstopdf}
\usepackage{color}

\begin{document}


\title{Laser-based acceleration of non-relativistic electrons at a dielectric structure}


\author{John Breuer}
\affiliation{Max Planck Institute of Quantum Optics, Hans-Kopfermann-Str.~1, 85748 Garching, Germany}
\author{Peter Hommelhoff}
\email[]{peter.hommelhoff@fau.de}
\affiliation{Department of Physics, Friedrich Alexander University Erlangen-Nuremberg, Staudtstr.~1, 91058 Erlangen, Germany}
\affiliation{Max Planck Institute of Quantum Optics, Hans-Kopfermann-Str.~1, 85748 Garching, Germany}


\date{\today}

\begin{abstract}
A proof-of-principle experiment demonstrating dielectric laser acceleration of \textit{non-relativistic} electrons in the vicinity of a fused-silica grating is reported. The grating structure is utilized to generate an electromagnetic surface wave that travels synchronously with and efficiently imparts momentum on 28\,keV electrons. 
We observe a maximum acceleration gradient of 25\,MeV/m. We investigate in detail the parameter dependencies and find excellent agreement with numerical simulations. With the availability of compact and efficient fiber laser technology, these findings may pave the way towards an all-optical compact particle accelerator. 
This work also represents the demonstration of the inverse Smith-Purcell effect in the optical regime.
\end{abstract}



\maketitle



The acceleration gradients of linear accelerators are limited by breakdown phenomena at the accelerating structures under the influence of large surface fields. Today's accelerators, which are based on metal structures driven by radio frequency fields, operate at acceleration gradients of $\sim$20--50\,MeV/m. The upper limit in future radio frequency accelerators, such as the discussed CLIC and ILC, is $\sim$100\,MeV/m, given by the damage threshold of the metal surfaces \cite{Lilje2004,Spataro2011,Solyak2009}. At optical frequencies dielectric materials withstand roughly two orders of magnitude larger field amplitudes than metals \cite{lenzner1998}. Together with the large optical field strength attainable with short laser pulses, dielectric laser accelerators (DLAs) hence may support acceleration gradients in the multi-GeV/m range \cite{Rosenzweig1995}. With this technology lab-size accelerators, providing particle beams with energies currently only available at km-long facilities, seem feasible. Here we demonstrate the efficacy of the concept.

Charged particle acceleration with oscillating fields requires an electromagnetic wave with a phase speed equal to the particle's velocity and an electric field component parallel to the particle's trajectory. So far, laser-based particle acceleration schemes employ the longitudinal electric field component of a plasma wave \cite{Leemans2006,Osterhoff2008,Lundh2011} or of a tightly focused laser beam \cite{Payeur2012}, but in both schemes the accelerating mode has a phase velocity that does not match the speed of light. Therefore, relativistic particles can only be accelerated over short distances and the maximum attainable energies of these devices are limited. Exploiting the near-field of periodic structures, for example of optical gratings, offers the possibility to continuously accelerate non-relativistic as well as relativistic particles. In essence, the effect of the grating is to rectify the oscillating field in the frame co-moving with the electron, conceptually similar to conventional radio frequency devices. Single gratings can only be used to accelerate non-relativistic electrons, an effect also known as the inverse Smith-Purcell effect \cite{Takeda1968,Mizuno1975,Mizuno1987}. However, double grating structures, in which electrons propagate in a channel between two gratings facing each other, support a longitudinal, accelerating speed-of-light eigenmode that can be used to accelerate relativistic particles \cite{plettner2006,Peralta2013}. 



Besides offering orders of magnitude larger acceleration gradients than conventional radio frequency accelerators, grating-based DLAs have the additional advantage of being scalable by linearly concatenating multiple gratings alongside the electron beam in a modular manner. The direct compatibility of non-relativistic and relativistic photonic grating structures enables an all-optical DLA. The single grating structures, presented here, can serve as a means to bridge the gap between the electron source and the relativistic DLA structures.

\begin{figure}
	\centering
		\includegraphics[scale=1]{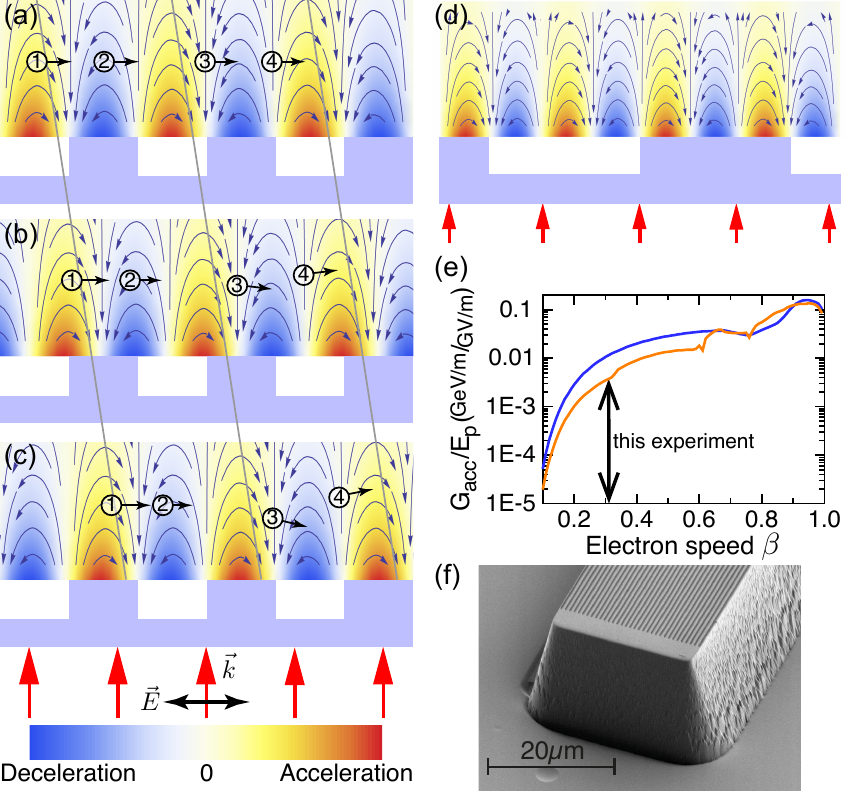}
	\caption{(a-c) Three subsequent conceptual pictures of four electrons (encircled numbers) passing the transparent grating (light blue structure, white: vacuum) that is illuminated by the laser from below. Time step between each picture: 1/4 optical period. The laser is linearly polarized in the plane of projection and propagates upward. Snapshots of the electric field distribution of the first spatial harmonic ($n=1$) are shown above the grating. It falls off exponentially with increasing distance from the grating (color coded) and co-propagates synchronously with the electrons along the grating surface. Depending on the position of the electron inside the field, the force acts accelerating (1), decelerating (2) or deflecting (3,4) as indicated by the blue arrows. (d) Electric field distribution of the third spatial harmonic ($n=3$) as used in this work. (e) Simulated ratio between the acceleration gradient $G_{\mathrm{acc}}$ and the applied laser peak electric field $E_{\mathrm{p}}$ for the first (blue) and third (orange) spatial harmonic for electrons passing a grating at a distance of 50\,nm. See text for details. (f) Scanning electron microscope image of the fused silica grating. 
	}
	\label{fig:grating}
\end{figure}

Acceleration at a grating is based on evanescent modes, known as spatial harmonics \cite{Plettner2011}. They are excited by laser light that impinges perpendicularly to the grating surface (Fig.\ \ref{fig:grating}) and consist of an electromagnetic wave that propagates parallel to the grating surface. The $n$-th spatial harmonic oscillates $n$ times per grating period ($n=1$ in Fig.\ \ref{fig:grating}(a-c), $n=3$ in Fig.\ \ref{fig:grating}(d)) and travels with a phase velocity $v_{\mathrm{ph}}=f \lambda_{\mathrm{p}}/n=c \lambda_{\mathrm{p}}/(n \lambda)$. Here, $\lambda_{\mathrm{p}}$  is the grating period, $\lambda$ the wavelength of the incident light, $f$ its frequency and $c$ the speed of light. Hence, the $n$-th harmonic is synchronous with electrons with the velocity $v=\beta c=v_{\mathrm{ph}}$, yielding the synchronicity condition $\beta=\lambda_{\mathrm{p}}/(n \lambda)$ \cite{plettner2009}. The field strength falls off exponentially with increasing distance from the grating surface with a decay constant $\Gamma=\beta \gamma \lambda/(2\pi)$, with $\gamma=1/\sqrt{1-\beta^2}$ \cite{Plettner2011}. 
Depending on the position of the electrons inside the laser field, the electric field vector acts accelerating, decelerating or deflecting (Fig.\ \ref{fig:grating}). The Lorentz force caused by other spatial harmonics that do not satisfy the synchronicity condition averages to zero over time. 

In our experiment we focus laser pulses derived from a long-cavity Titanium:sapphire os\-cil\-la\-tor \cite{Naumov2005} onto a fused silica transmission grating. The laser beam parameters are: center wavelength of $\lambda= 787$\,nm, repetition rate of $f_{\mathrm{rep}}= 2.7$\,MHz, pulse duration of $\tau_{\mathrm{p}}=110$\,fs and focal waist radius of $w_{\mathrm{l}}=(9\pm0.4)\,\mu$m. We chose the grating period to $\lambda_{\mathrm{p}}=750$\,nm, hence the third spatial harmonic ($n=3$) is synchronous with 27.9\,keV electrons and decays with $\Gamma=42$\,nm.
This choice of $\lambda_\mathrm{p}$ was a trade-off between a lower bound set by the grating manufacturer 
and the decreasing excitation efficiency for higher spatial harmonics (Fig.\ \ref{fig:grating}(e)). 

\begin{figure}
	\centering
		\includegraphics[scale=1]{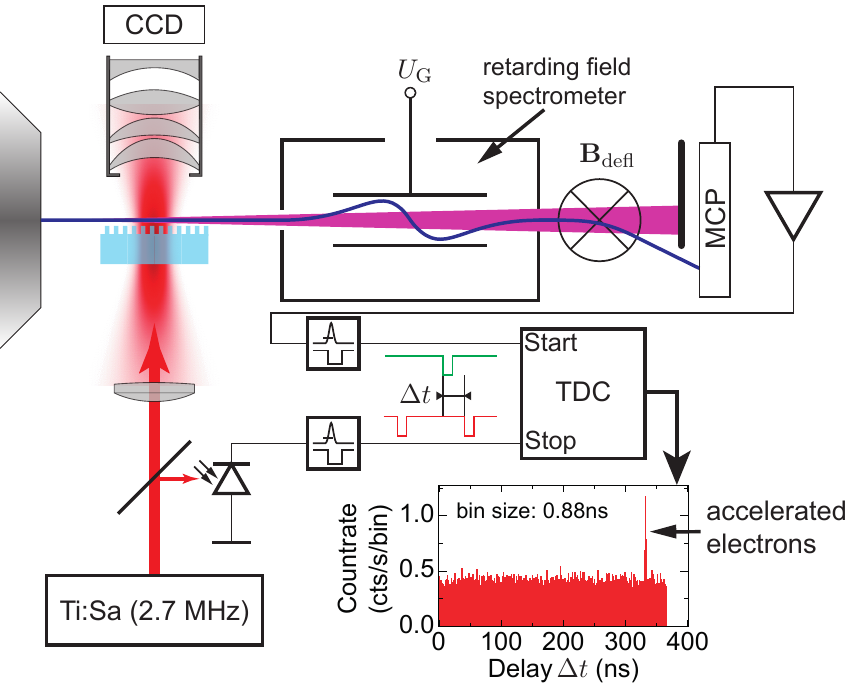}
	\caption{Sketch of the experimental setup and detection scheme. Electrons emitted from a scanning electron microscope column (left) pass the transparent grating (light blue structure). Here they interact with the accelerating field excited by the laser pulses (red, propagating from bottom to top). A microscope objective is used to monitor the position of the laser focus. The electrons that can pass the spectrometer (center) are detected at the MCP. 
	}
	\label{fig:setup}
\end{figure}

Fig.\ \ref{fig:setup} depicts the scheme of the experimental setup. We use the column of a conventional scanning electron microscope (SEM) as electron source that provides a DC electron beam with a 1/$e$ focal waist radius of $w_{\mathrm{e}}=(70\pm20)$\,nm 
and a beam current of $I_{\mathrm{b}}=(4.2\pm0.5)$\,pA. Because of the continuous-wave nature of the electron beam only a  small fraction of electrons interacts with the laser pulses ($I_{\mathrm{eff}}=I_{\mathrm{b}} \tau_{\mathrm{p}} f_{\mathrm{rep}}\approx 10$ electrons per second). However, the excellent beam control of a standard SEM outweighs the low expected count rate in this proof-of-concept experiment. With a pulsed electron source, which will be implemented in future experiments, $I_{\mathrm{eff}}$ will be many orders of magnitude higher; 3\,A peak current have been shown with a pulsed source, 10$^{12}$ times as large as $I_\mathrm{b}$ in our experiment \cite{Ganter2008}. After passing the grating, the electrons enter an electrostatic filter lens spectrometer \cite{brack1962}, which blocks all unaccelerated electrons. 
A micro-channel plate detector (MCP) is used to detect the electrons that pass the filter lens. The current of accelerated electrons $I_{\mathrm{acc}}$ is recorded with a digital lock-in scheme, with which we measure the time delay between detector events and laser pulses. The correlated signal of accelerated electrons appears as a peak at a fixed delay, whereas background counts ($\sim$50-70\,cts/s) are uniformly distributed over all delays. 


\begin{figure}
	\centering
		\includegraphics[scale=1]{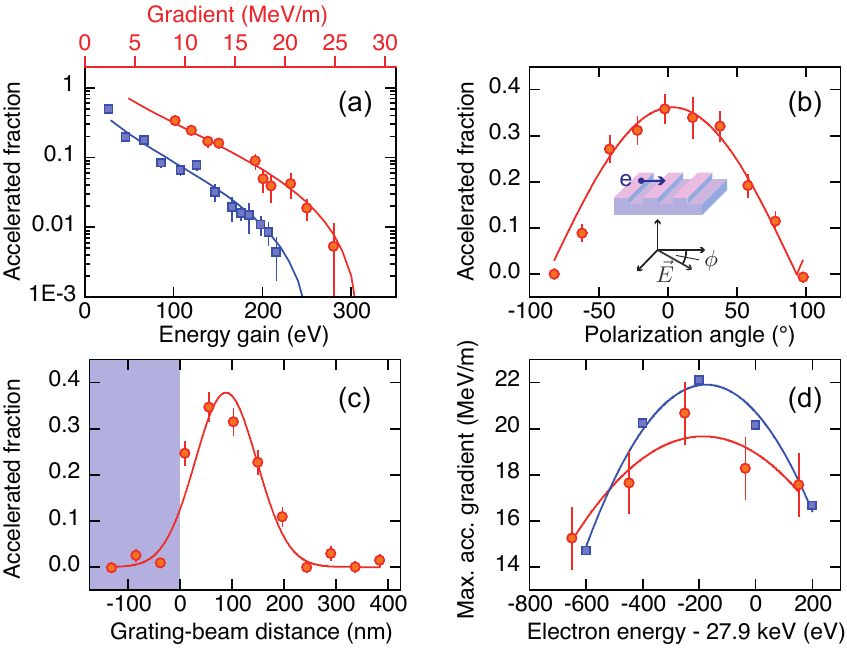}
	\caption{(a) Measurement of the accelerated fraction of electrons $I_{\mathrm{acc}}/I_{\mathrm{eff}}$ as a function of energy gain (bottom axis) and acceleration gradient (top axis), for two different laser peak electric fields [$E_{\mathrm{p}}=2.85$\,GV/m (orange circles), $E_{\mathrm{p}}=2.36$\,GV/m (blue squares)], with corresponding simulated curves. The only free parameters of the simulated curves are the distance $z_0$ of the electron beam center from the grating surface, implying $z_0=(120\pm10)$\,nm, and the overall amplitude. (b) Dependence of the accelerated fraction on the laser polarization angle $\phi$. The data can be well fitted with the expected sinusoidal behavior (orange curve). 0$^{\circ}$ means that the laser polarization is parallel to the electrons' momentum, 90$^{\circ}$ that it is perpendicular to it (see inset). (c) Measurement of the accelerated fraction versus the relative distance between the grating surface (shaded area) and the electron beam center. Note that due to the finite width of the electron beam some electrons are accelerated although the beam center lies inside the grating. Solid line: Gaussian fit. (d) Measurement (orange circles) and ab initio simulation (blue squares) of the maximum acceleration gradient as a function of the initial electron energy. This measurement confirms that efficient acceleration occurs only if the synchronicity condition is satisfied. The lines are guides to the eye. 
	}
	\label{fig:results}
\end{figure}

We now discuss detailed measurements confirming the observation of dielectric laser acceleration of non-relativistic electrons. In Fig.\ \ref{fig:results}(a)  we show the accelerated fraction as a function of the energy gain $\Delta E$ for two different laser peak electric fields of 2.36\,GV/m and 2.85\,GV/m. The measured accelerated fraction equals the ratio of the integrated number of accelerated electrons $I_{\mathrm{acc}}$ that gain more energy than $\Delta E$ to the number of electrons $I_{\mathrm{eff}}$ that can interact with the laser pulses. 
The acceleration gradient equals $G_{\mathrm{acc}}=\Delta E/x_{\mathrm{acc}}$, with the effective acceleration distance $x_{\mathrm{acc}}=11.2\,\mu$m, which we obtain from $\beta$, $w_{\mathrm{l}}$, $\tau_{\mathrm{p}}$ and the Gaussian shaped laser pulse. We measure a maximum acceleration gradient of 25.0\,MeV/m.

An eigenmode method \cite{Pai1991} in conjunction with particle tracking in the resulting fields allows us to obtain the accelerated fraction of electrons numerically. The results show perfect agreement with the experimental data (Fig.\ \ref{fig:results}(a)), assuming a Gaussian electron beam profile with its center $z_0=(120\pm10)$\,nm away from the grating surface and a 1/$e$ beam radius $w_{\mathrm{e}}=77$\,nm, matched to the experiment. We further infer from our simulations that the maximum acceleration gradient of 25\,MeV/m results from electrons that, due the finite beam width, pass the grating at a distance of $\sim$50\,nm. We have tried to position the beam closer to the grating surface, but beam clipping and residual surface charging presumably prevented a closer approach in the current setup.

The electrons are only accelerated by the electric field component that is parallel to their momentum, as can be seen from the dependence of the acceleration on the polarization of the laser electric field, in Fig.\ \ref{fig:results}(b). This strongly supports acceleration with the electromagnetic light \textit{field} and clearly rules out the much weaker intensity-dependent but polarization-independent ponderomotive acceleration \cite{Boot1957}, whose effect we estimate to a vanishing $\sim$10\,keV/m for the given laser parameters.

In Fig.\ \ref{fig:results}(c) we show the accelerated fraction as a function of distance from the grating surface. The data can be fitted with a Gaussian of width $(119\pm11)$\,nm and agrees well with simulations of the accelerated fraction assuming a 1/$e$ electron beam radius of 77\,nm. This lies right within the experimentally obtained $w_{\mathrm{e}}=(70\pm20)$\,nm. The fact that acceleration is only possible in the vicinity of the grating surface confirms that the accelerating fields fall off steeply away from the grating, as expected for this near-field-based acceleration scheme.

In Fig.\ \ref{fig:results}(d) we present a measurement of the maximum acceleration gradient as a function of the initial electron energy, verifying that efficient acceleration only occurs if the electron velocity is matched to the grating period, according to the synchronicity condition. We observe maximum acceleration for an initial energy of 27.7\,keV. The ab initio calculations in Fig.\ \ref{fig:results}(d) represent the maximum acceleration gradient as function of electron energy 60\,nm away from the grating surface and show excellent agreement with the experiment.

In the current setup, the peak laser field $E_{\mathrm{p}}$ is laser-power-limited to 2.85\,GV/m. Damage threshold measurements of fused silica gratings \cite{Soong2011b} indicate that $E_{\mathrm{p}}$ can be increased by a factor of $\sim$3.4, 
up to 9.4\,GV/m. With this $E_{\mathrm{p}}$, Fig.\ \ref{fig:grating}(e) shows a simulation of the acceleration efficiency, that is, the ratio between the maximum acceleration gradient and the laser peak field $E_\mathrm{p}$. From this we infer that electrons with $\beta=0.3$ (25\,keV) passing the grating surface at 50\,nm become accelerated with 350\,MeV/m, assuming synchronicity with the first spatial harmonic. Exploiting the third spatial harmonic like in this letter reduces the achievable acceleration gradient by a factor of three to 110\,MeV/m. Furthermore, the acceleration efficiency increases steeply for relativistic electrons. Therefore, we expect that electrons with $\beta=0.95$ (1.1\,MeV) experience an acceleration gradient as large as 1.7\,GeV/m for synchronicity with the first spatial harmonic.


\begin{figure}
	\centering
		\includegraphics[scale=1]{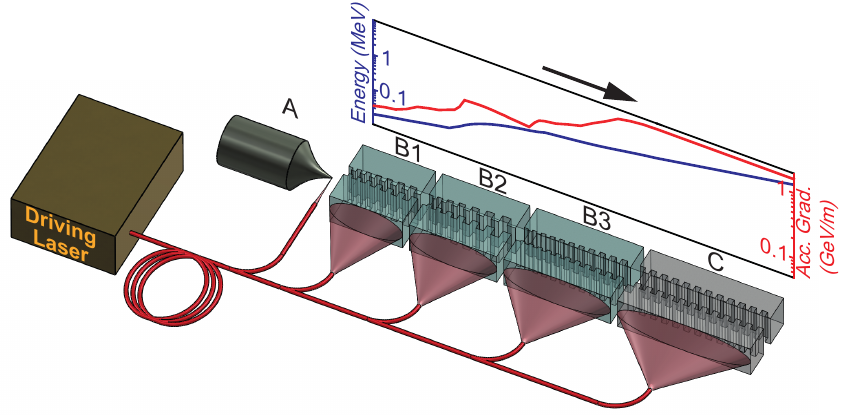}
	\caption{Sketch of an envisioned layout of an all-optical linear accelerator with electron gun (A), non-relativistic (B) and relativistic sections (C), all driven by a common, for ease of operation fiber-based, laser source. The non-relativistic section consists of grating structures with a tapered grating period to assure synchronicity with the accelerating electrons using the third (B1), second (B2) and first (B3) spatial harmonic. Inside the non-relativistic section, the channel width can increase because of the increasing decay constant $\Gamma$ of the accelerating fields. Not to scale.}
	\label{fig:DLA}
\end{figure}

The intriguing feature of non-relativistic DLA structures is the inter-compatibility with their relativistic counterparts. Thus, with the demonstrations of both relativistic \cite{Peralta2013} and non-relativistic dielectric laser acceleration, operating with identical laser sources at the first and third spatial harmonic, respectively, an all-optical dielectric-based linear accelerator can now be seriously considered. Corroborating this, we note that based on double grating structures all components of a conventional accelerator, such as deflecting, focusing and bunching structures can be realized \cite{plettner2009,Plettner2011}. 

A schematic of the envisioned DLA design is shown in Fig.\ \ref{fig:DLA}. Due to their micron-scale size, DLA structures require and support electron beams with normalized emittance values in the nm-range \cite{Cowan2008}. Therefore, a non-relativistic DLA section in combination with 
an ultralow emittance electron source, such as a laser-triggered \mbox{needle} cathode with an intrinsic emittance of $\sim$50\,nm \cite{Ganter2008}, 
is crucial for the injection of high-quality electron beams into the relativistic section of the DLA. 
A common laser source, amplified in various sections, can be used to drive the electron gun, the non-relativistic and the relativistic DLA. Here, fast progress in fiber laser technology, especially phase-coherent splitting and amplification may play an important role, similar to what has been proposed for laser-plasma-based acceleration schemes \cite{Mourou2013}. This greatly facilitates the synchronization and optical phase-stability between the different acceleration stages, which is essential for proper functioning
. The non-relativistic part (up to $\sim$1\,MeV) may consist of subsections of tapered dielectric gratings in which an adaptively increasing grating period accounts for the change in electron velocity. 
For this part, either single or double grating structures may be used. While single gratings offer a simpler setup, the evanescent nature of the acceleration can lead to beam distortion. Double grating structures have the advantage of larger efficiency and a symmetric profile of the accelerating fields \cite{Plettner2011}. Exploiting different spatial harmonics allows to overcome fabrication limitations on the grating period by starting, for example, with 30\,keV electrons at the point of injection. After using the third spatial harmonic to accelerate up to $\sim$50\,keV, one may switch to the more efficient second harmonic and at $\sim$400\,keV to the first harmonic to accelerate further.

Because the electron beam quality 
is directly related to the brightness and the shortest achievable wavelength in X-ray free electron lasers (FELs), DLAs may find application in future low-cost, compact, high brilliance sources of hard X-rays, with the potential to open experiments in biology, medicine and materials science to a broad community of users \cite{Carlsten2010}. Despite the low bunch charge ($\sim$fC) supported by DLAs, successful FEL operation appears feasible \cite{plettner2008b}; source development is needed, as in other FEL approaches \cite{Ganter2008,Polyakov2013}. Moreover, the realization of accelerating, deflecting, focusing and bunching elements for \textit{non-relativistic} electrons could lead to a new generation of electron optics with applications in ultrafast electron diffraction experiments and time-resolved electron microscopy. Next experimental steps will comprise the implementation of concatenated, symmetric double-grat\-ing structures to demonstrate acceleration to several times the initial beam energy, as well as the combination with a laser-triggered electron source.

After writing of the manuscript we became aware of a recently proposed innovative scheme combining plasma-based acceleration with the periodic field reversal at grating structures, which may lead to scalable accelerators with a sustained acceleration gradient up to TeV/m \cite{Pukhov2013}.

\begin{acknowledgments}
We gratefully acknowledge R.\ Graf for support of laser system operation, A.\ Apolonski and F.\ Krausz for loan of the long-cavity oscillator, J.\ Hoffrogge for work on the electron column, H.\ Ramadas and R.\ Davies for simulation work, P.\ Altpeter for titanium coating and the Stanford DAPRA AXiS collaboration for discussions. This work has been funded by the Max Planck Society and Munich-Centre for Advanced Photonics.
\end{acknowledgments}

\bibliography{libscience}

\end{document}